\documentclass{aa}

\newcommand{\grs}    {GRS 1758$-$258}

\newcommand{\grp}    {${\rlap.}^{\circ}$}
\newcommand{\pri}    {${\rlap.}^{\prime \prime}$}
\newcommand{\rl}     {${\rlap.}^{s}$}

\newcommand{\ltsima} {$\; \buildrel < \over \sim \;$}
\newcommand{\simlt}  {\lower.5ex\hbox{\ltsima}}            
\newcommand{\gtsima} {$\; \buildrel > \over \sim \;$}
\newcommand{\simgt}  {\lower.5ex\hbox{\gtsima}}            

\usepackage{graphicx}
\usepackage{txfonts}
\begin{document}

   \title{Real-time evolution of a large-scale relativistic jet }


   \author{Josep Mart\'{\i}\inst{1,5}
          \and
          Pedro L. Luque-Escamilla\inst{2,5}
          \and
          Gustavo E. Romero\inst{3,4,5}
          \and
          Juan R. S\'anchez-Sutil\inst{5}
          \and
         \'Alvaro J. Mu\~noz-Arjonilla\inst{5} 
          }

\institute{Dept. de F\'{\i}sica, EPS de Ja\'en, Universidad de Ja\'en, Campus Las Lagunillas s/n, A3-402, 23071 Ja\'en, Spain\\
\email{jmarti@ujaen.es}
\and
Dept. de Ingenier\'{\i}a Mec\'anica y Minera, EPS de Ja\'en,  Universidad de Ja\'en, Campus Las Lagunillas s/n, A3-402, 23071 Ja\'en, Spain\\
\email{peter@ujaen.es}
\and
Instituto Argentino de Radioastronom\'{\i}a, C.C.5, 1894 Villa Elisa, Buenos Aires, Argentina
\and
Facultad de Ciencias Astron\'omicas y Geof\'{\i}sicas, Universidad Nacional de La Plata, Paseo del Bosque S/N, 1900, La Plata, Argentina\\
\email{gustavo.esteban.romero@gmail.com}
\and
Grupo de Investigaci\'on FQM-322, Universidad de Ja\'en, Campus Las Lagunillas s/n, A3-065, 23071 Ja\'en, Spain\\
\email{jrssutil@ujaen.es, ajmunoz@ujaen.es}
}

   \date{Received xxxx, 2015; accepted xxxx, xxxx}

 
  \abstract   
    {Astrophysical jets are ubiquitous in the Universe on all scales, but their large-scale dynamics and evolution in time
     are hard to observe since they usually develop at a very slow pace.}
   {We aim to obtain the first observational proof of the expected large-scale evolution and interaction with the environment in an astrophysical jet.
  Only jets from microquasars offer a chance to witness the real-time, full-jet evolution within a human lifetime, 
  since they combine a $^\backprime$short$^\prime$, few parsec length with relativistic velocities.}
  {The methodology of this work is based on a systematic recalibraton of interferometric radio observations of microquasars 
   available in public archives. In particular, radio observations of the microquasar \grs\ over  less than two decades have provided
  the most striking results.}
   {Significant morphological variations in the extended jet structure of  \grs\ are reported here that were previously missed.
   Its northern radio lobe underwent a major morphological variation that rendered the hotspot undetectable in 2001 and reappeared again in the following years.
  The reported changes confirm the Galactic nature of the source. We tentatively interpret them  in terms of the growth of instabilities in the jet flow.
  There is also evidence of surrounding cocoon. These results can provide a testbed for models accounting for 
  the evolution of jets and their interaction with the environment.}
   {}

  
   \maketitle
%
 
\section{Introduction}

Astrophysical jets are observed in a variety of environments \citep{2005AdSpR..35..908D} including young stellar and Herbig-Haro (HH) objects 
\citep{2001ARA&A..39..403R}, planetary nebulae \citep{1998AJ....116.1357S}, 
microquasars \citep{1999ARA&A..37..409M}, active galactic nuclei (AGN), and distant quasars \citep{1984RvMP...56..255B}. 
These outflows are triggered when the magnetic field taps the rotational energy from a central compact object or disk \citep{1982MNRAS.199..883B, 89879520150101}. 
This mechanism is characterized well because the mass accretion and ejection events evolve fast enough to be appropriately 
sampled with multi-wavelength observations of both microquasars \citep{1998A&A...330L...9M, 1994Natur.371...46M, 1999MNRAS.304..865F}
and AGN  \citep{2000Sci...289.2317G, 2002Natur.417..625M}, where these processes occur on time scales from hours or days to years, respectively. 
However, the long-term evolution of the jet flow far away from the central engine is not understood as well.
The large sizes of relativistic jets in AGN make them evolve on characteristic times
of $\sim10^6$ yr  \citep{2008MNRAS.388..625S}, while the low velocities of HH jets imply shorter characteristic times  of $\sim 10^3$ yr \citep{2001ApJ...551..347D}. 
Extended jets in both types of systems will then appear almost frozen during a human lifetime. 

In this context, only microquasar jets that combine relatively short lengths and relativistic velocities \citep{1999ARA&A..37..409M}
 offer a chance to study large-scale jet dynamics and its interaction with the environment in almost real time. Here, we have payed attention to  \object{\grs}\  which
is believed to be a microquasar associated with a low-mass X-ray binary system. 
Although its Galactic nature has never been confirmed because of the lack of optical and infrared spectra of the companion star \citep{2014ApJ...797L...1L},
\grs\ appears as a very bright and persistent hard X-ray source  towards the Galactic centre region \citep{1991A&A...247L..29S}. 
This source has strong spectral similarities in X-rays with the classical black hole candidate \object{Cygnus X-1} \citep{1999ApJ...525..901M}.
It displays double-sided radio jets \citep{1992ApJ...401L..15R,1994AIPC..304..413M} whose arc-minute extension implies a parsec-scale linear size 
if located at the distance of the Galactic centre, hereafter assumed to be 8.5 kpc.
In this Letter, we revisit the huge archive of \grs\ observations at radio wavelengths obtained with the Very Large Array (VLA) interferometer 
of the National Radio Astronomy Observatory (NRAO). This provides us with a unique set of highly sensitive maps with an angular resolution that is
well suited to exploring the evolution of the \grs\ jets over more than a decade.

\section{Data analysis and results}

We data-mined the public archives of the VLA interferometer hosted by NRAO. 
The angular size and morphology of the arcminute \grs\  jets are well sampled by using the 6 cm wavelength and the C-configuration of the array. 
Two intermediate frequency bands 50 MHz wide were available from the 
VLA correlator products. A total of four useful observing projects between 1992 and 2008 could be retrieved with this instrumental setup (see first block of Table \ref{log}). 

 All downloaded projects were individually recalibrated by using the AIPS software package of NRAO, 
  taking special care to remove corrupt visibilities in both the target and the calibrator sources. 
  The phase calibrator was \object{J1751$-$253}, while the visibility amplitude was tied to the known flux densities of \object{3C286} and \object{3C48}. 
  Some older projects needed to be transformed from the B1950.0 to the J2000.0 reference system, 
  by using the AIPS task UVFIX, in order to make the multi-epoch map comparison easier.
Radio maps were created and deconvolved using the CLEAN algorithm as provided within the IMAGR task of AIPS. 
Their respective synthesized beams were individually determined first and later averaged. 
The resulting averaged elliptical Gaussian beam was finally used to convolve the clean components of each observing epoch in a second run of the IMAGR task. 

\begin{figure*}
   \centering
   \includegraphics[angle=0,width=16.0cm]{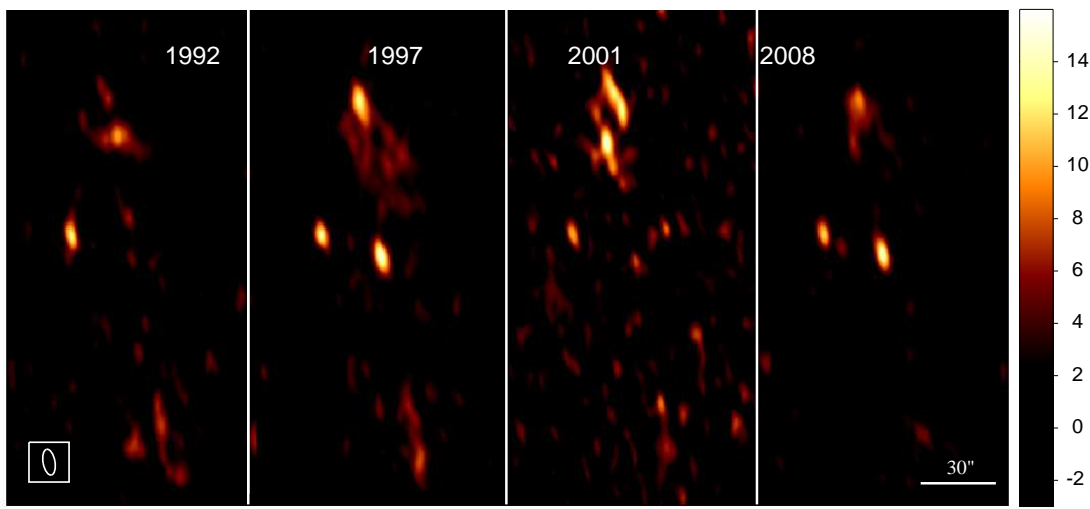}
      \caption{Time evolution of the \grs\ extended radio jets as observed with the VLA interferometer at the 6 cm wavelength (4.8 GHz) 
      over sixteen years (1992-2008) with nearly identical angular resolution. 
     North is up and east to the left. The horizontal bar at the bottom right corner shows the angular scale. 
     The  interferometric synthesized beam is 
      10\pri 50 $\times$ 4\pri 75, with position angle of  $10^{\circ}$ (bottom left ellipse).
      The vertical colour bar provides a linear brightness scale in units of $\mu$Jy beam$^{-1}$.  
      The rms background noise is 10, 10, 19 and 8 $\mu$Jy beam$^{-1}$ for the 1992 to 2008 frames, respectively. 
      The two variable point sources are  the \grs\ 
      central core and an unrelated object about $25^{\prime\prime}$ to the east of it.
    }
         \label{sequence}
   \end{figure*}

In Fig. \ref{sequence}, we present the final sequence of 6 cm maps showing the appearance of the \grs\ during the years 1992, 1997, 2001, and 2008 on arcminute scales. 
The different panels can be considered as approximate matching-beam maps with similar point spread functions. 
Having the same angular resolution, they are therefore suitable for meaningful comparison of the jet morphological variations visible in different epochs.
 In all cases, natural weighting of the interferometric visibilities (i.e. a +5 value of the IMAGR ROBUST parameter) was applied to maximize sensitivity to extended emission. 
 The AIPS task DBCON was used to concatenate short time slots of different VLA monitoring projects into a single observing epoch. 


In Table \ref{obsdata} we compile the positions, deconvolved angular sizes, flux densities coming from the northern hotspot component that dominates the lobe emission, 
and its radio luminosities in the 0.1-$10^5$ GHz range, assuming a $-0.7$ typical spectral index for non-thermal radio emission. 
We additionally include the minimum energy content and magnetic field assuming equipartition of energy between relativistic particles and magnetic field, 
using the \citet{pacholczyk1970radio} formulation. An estimate of the relativistic electron density needed to account for the observed radio luminosity is also given.

Finally, a few additional archive projects exist in the C, D and CD configurations  (see 2nd block of Table \ref{log}). 
Although there was not high enough quality to provide new individual frames in Fig. \ref{sequence}, 
they were selected to enhance the sensitivity to extended emission in deep imaging that is discussed below.

\begin{table}
\caption{Log of VLA  6 cm observations used in this work.}          
\label{log}      
\centering                        
\begin{tabular}{c c l c c}     
\hline\hline                
Project    &           VLA             &  Observation     &   On-source       &      Central     \\
 code      &   config.       &       ~~~~~date            &    time (s)          &  Julian day  \\
\hline
AM345        &         C       &         1992 Mar 21        &      1770     &                                     \\
                   &                   &        1992 Apr 09           &    6040       &       2448719     \\
                   &                  &         1992 Apr 11           &    2610       &                          \\
AM560         &       C        &        1997 Aug 03           &   2440     &                      \\
                     &                 &        1997 Aug 05            &  2440     &  \\
                     &                  &       1997 Aug 08           &   2440 &   \\
                     &                  &      1997 Aug 11           &   2160   &  \\  
                      &                &        1997 Aug 14            &  2170   &           2450674  \\
                      &                 &       1997 Aug 15             & 1860 &   \\
                      &                 &     1997 Aug 18         &     2450  &  \\
                      &                  &   1997 Aug 20         &    2430  &  \\
                     &                 &        1997 Aug 24          &    2420      &  \\
AR458            &    C         &       2001 Jul 08             &     473 &  \\
                       &                 &      2001 Jul 24              &    533   &  \\
                       &                &      2001 Aug 4                &  673   &  \\
                       &                 &      2001 Aug 9              &    603 &  \\
                       &               &        2001 Aug 16            &    453  &            2452132 \\
                       &               &       2001 Aug 26               & 413  &   \\
                        &               &       2001 Aug 30             &   403  &  \\
                        &               &       2001 Sep 07             &    533     &  \\
AS930           &     C          &      2008 Apr 01         &      4840 &   \\
                      &                 &      2008 Apr 07          &     4840  &            2454563 \\
                     &                   &      2008 Apr 12          &     4840  &   \\          
\hline
AM345          &      D         &      1992 Sep 26-27     &    5690     &         2448892    \\
AM428          &     CD        &      1993 Oct 3-4          &   6460      &        2448899 \\
AR476          &      C          &       2002 Oct 15          &    1453     &   \\
                    &                    &       2002 Oct 16           &   1513     &         2452582  \\
                     &                   &       2002 Nov  11             &  1623   &    \\
                      &                  &       2002 Dec 2             &   1533 &   \\
AR523          &     C            &      2004 Apr 30            &    953      &        2453126   \\
                      &                  &       2004 May 05            &   953       &       2453131 \\
\hline
\end{tabular}
\end{table}

 \section{Discussion}

    In 1992, the jet's northern lobe in Fig. \ref{sequence} ended in a sort of bow-shaped working surface with a conspicuous component at its vertex that we interpret as the terminal hotspot. 
   Comparison with the 1997 frame, where the hotspot is brighter and easily recognized, implies a noticeable shift of $13^{\prime\prime} \pm 1^{\prime\prime}$ on a time interval of 5.4 yr.
   This value is consistent with previous, very conservative upper limits that supported the idea of a continuously powered stationary jet flow 
   \citep{2002A&A...386..571M,2011MNRAS.415..410S}.
    When assuming the hotspot identification is correct,  the 2\pri 4 $\pm$ 0\pri 2 yr$^{-1}$ associated proper motion translates into a velocity of the projected jet head of ($0.32 \pm 0.03)c$, 
    where $c$ is the speed of light. In the 2001 frame, the northern jet lobe was observed to evolve into two elongated fragments that were nearly parallel to the jet direction. 
    The conspicuous hotspot of 1997 completely lost its shape, and only traces of its extended emission remained at its position after four years. 
    Seven years later, in 2008, a new hotspot had clearly reappeared at some time in between. 
    Had the 2008 northern hotspot existed in 2001, it should have been detected well above the rms noise. 
    Moreover, from Table \ref{obsdata}
   the position offset between the 1997 and 2008 hotspots is estimated to be 1\pri 9 $\pm$ 0\pri 7. Thus, we are observing a newly formed structure. 
   On the other hand, the southern counter-jet is also visible in Fig. \ref{sequence}, but it did not display a structure that was as well organized as the northern jet.
   Only hints of precession previously noticed by \cite{2002A&A...386..571M}  are evident in 1997, but they are remarkably absent in the even more sensitive 2008 frame. 
   This is additional evidence that the \grs\ jet evolution is real.

\begin{figure}
   \centering
   \includegraphics[angle=0,width=10.0cm]{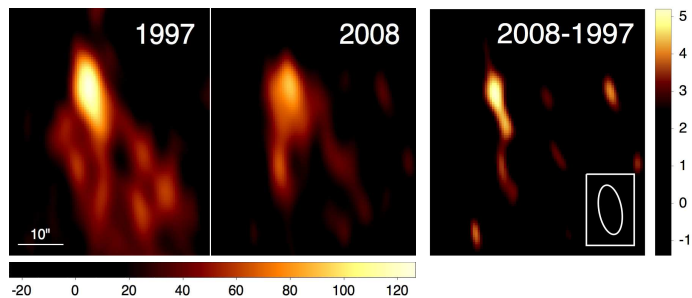}
      \caption{
      Close up of the northern radio lobe of \grs\ as observed with the VLA at 6 cm in 1997 (left) and 2008 (middle), when the hotspot was fainter. 
      The coloured horizontal bar  provides a linear intensity scale in $\mu$Jy beam$^{-1}$.
       The bottom left corner bar  gives the angular scale. North is up and east left. 
       The right panel shows the residual difference between the two epochs.
       Its colour vertical bar is scaled in units of the rms noise in the difference map (10 $\mu$Jy beam$^{-1}$).
       The synthesized beam in all panels is the same as in Fig.\ref{sequence} (bottom right ellipse).
              }
         \label{zoom_hotspot}
   \end{figure}

 The low declination of the source limits the achievable dynamic ranges of the Fig. \ref{sequence} maps. 
Therefore, difference maps were computed to assess the reliability of the observed structural variations better.
Our two more sensitive observing epochs were used for this purpose.  Figure \ref{zoom_hotspot} shows a zoomed view of morphology differences in the northern hotspot, 
together with  the subtraction of the 1997 clean component model from the 2008 visibility data.
 Residuals emerge at the $4\sigma$ to $5\sigma$ level. Beyond the field of view shown here,
they are also aligned along the jet position angle, suggesting changes that affect the whole jet flow.

The first consequence of the observed phenomena is an upper limit to the \grs\  distance. This comes from the time scale ($\tau \sim 11$ yr) of major structural changes.  
The northern jet flow was fully renewed between 1997 and 2008, as shown in Figs. \ref{sequence}, and \ref{zoom_hotspot}. 
Given its angular size ($\theta \sim 60^{\prime\prime}$), causality arguments dictate that the maximum possible distance lies within the boundaries of the Milky Way 
($\sim c \tau / \theta \sim 12$ kpc). Having the jet inclined with respect to the line-of-sight would  reduce this value further.
Therefore, the Galactic origin of \grs\  becomes fully confirmed for the first time.

 \begin{table*} 
 \caption{Observational parameters for the northern hotspot of \grs\ at the 6 cm wavelength }
 \label{obsdata}    
\centering                        
\begin{tabular}{ccccccccc}       
\hline\hline  
Epoch  &      R. A.  &  Dec.      &         Flux        &      Deconvolved     &       Radio                          & Minimum   &    Magnetic   &    Relativistic  \\
             &      (J2000.0) & (J2000.0)    &      density      &     angular size        &  luminosity                &   energy      &       field       &      e- density \\
             &        $18^h 01^m$  & $-25^{\circ} 43^{\prime}$             &      (mJy)        &     (arc-second$^2$)      &    (erg s$^{-1}$)          & (erg)           &  (Gauss)      &       (cm$^{-3}$)    \\ 
\hline 
  1992   &   12\rl 8 $\pm$ 0\rl 1 &    $48^{\prime\prime} \pm   1^{\prime\prime}$    &  $0.37 \pm 0.03$  &       $ (18 \pm 3) \times ( \leq 9 )$     &  $1.0 \times 10^{31}$ &   $\leq  3.3 \times 10^{44}$  &  $\geq 4.0 \times 10^{-5}$ &   $ \geq 7.6 \times 10^{-4}$ \\
\hline       
   1997   &  13\rl 00 $\pm$  0\rl 02  &  35\pri 6 $\pm$  0\pri 7   & $ 0.26 \pm  0.02$  &      $ (15 \pm 2) \times (6 \pm 1)$    &  $7.1 \times 10^{30}$  &    $1.9 \times 10^{44}$        &  $ 4.7 \times 10^{-5}$  &  $ 1.0 \times 10^{-3}$   \\
 \hline  
  2001$^a$  &   12\rl 9 $\pm$ 0\rl 1 & $33^{\prime\prime} \pm  1^{\prime\prime}$  &      $ 0.29 \pm 0.04$   &     $ (25 \pm 4) \times (4 \pm 2)$  &  $ 7.9 \times 10^{30}$  &      $ 2.1 \times 10^{44}$   &      $ 4.6 \times 10^{-5}$  &      $ 1.2 \times 10^{-3}$  \\
                    &  13\rl 13 $\pm$ 0\rl 03 & $52^{\prime\prime}  \pm 1^{\prime\prime}$ &       $ 0.26 \pm 0.04$  &   $(24 \pm 6) \times (3 \pm 2)$   &  $ 7.1 \times 10^{30}$   &       $ 1.6 \times 10^{44}$  &    $ 5.1 \times 10^{-5}$  &   $ 1.1 \times 10^{-3}$ \\ 
\hline
  2008         &  13\rl 11s $\pm$ 0\rl 03 & 36\pri  8  $\pm$  0\pri 6  &          $0.18 \pm 0.02$  &  $(13 \pm 2) \times (8 \pm 1)$ &     $ 4.9 \times 10^{30}$  &  $ 1.6 \times 10^{44}  $  &  $ 3.9 \times 10^{-5}$  &  $ 7.4 \times 10^{-4}$ \\
 \hline
 \end{tabular}
 \tablefoottext{a}{Northern hotspot appears to have fragmented into two components. Data for both are given.}
 \end{table*}

The physical interpretation of these morphological changes is uncertain, 
although their disruptive appearance suggests that they may be due to the onset of hydrodynamic instabilities 
\citep{1997atas.conf...17B} such as Kelvin-Helmholtz (KH) or Rayleigh-Taylor (RT) instabilities. 
The stability condition essentially depends on the ratio $\eta = n_j / n_a$ with $n_j$ and $n_a$ the density of the jet and the ambient medium, respectively. 
If $\eta < 1$, the jet may be unstable. 
When assuming that \grs\ jet is baryonic and that most of the mass flux is in the form of a thermal plasma (e.g., about 90\%), 
the relativistic particles responsible for the non-thermal radio emission, whose density of $\sim 10^{-3}$ cm$^{-3}$ has been estimated from equipartition 
(see Table \ref{obsdata}), account for the remaining 10\%. 
Thus, the actual jet density is tentatively estimated as $n_j \sim 10^{-2}$ cm$^{-3}$. 

Given that microquasars are typically located in much less dense environments than the canonical interstellar medium \citep{2002A&A...388L..40H}, 
we assume $n_a \sim 0.1$ cm$^{-3}$. This yields a density contrast in the vicinity of the terminal hotspot of 
$\eta \sim 10^{-1}$. Therefore, the jet could be prone to undergoing
 hydrodynamic instabilities, and its disruption may occur if the KH or RT modes have enough time to grow up to a length scale 
comparable to the jet radius $r_j$. These time scales may be calculated \citep{2009A&A...503..673A,2010IJMPD..19..931A}
as $t_{KH} \sim  (2 r_j /c) \eta^{1/2}$ and $t_{RT} \sim  (2 r_j /c)(2\eta /3)^{1/2}$, respectively. 
When assuming that the filaments of radio emission converging to the hotspot in Fig. \ref{zoom_hotspot} are actually tracing the jet flow, 
$r_j \sim 0.1$ pc for \grs\  at its estimated distance. 
Taking this value, growth time scales of destructive RT and KH instabilities are found to be very similar and to last several months. 

Considering the uncertainties involved in this calculation, it thus appears conceivable from the physical point of view that the \grs\  jets undergo RT and/or KH instabilities 
that completely reorganize their collimated outflow on a yearly time scale, as suggested by the multi-epoch observations reported in this work. 
In the maps presented in Fig. \ref{sequence}, we have merged data sets scattered over no more than a two-months span,  so that
they are safely below the limit for avoiding an excessive smearing of the imaged structures.

\begin{figure}
   \centering
   \includegraphics[angle=0,width=7.5cm]{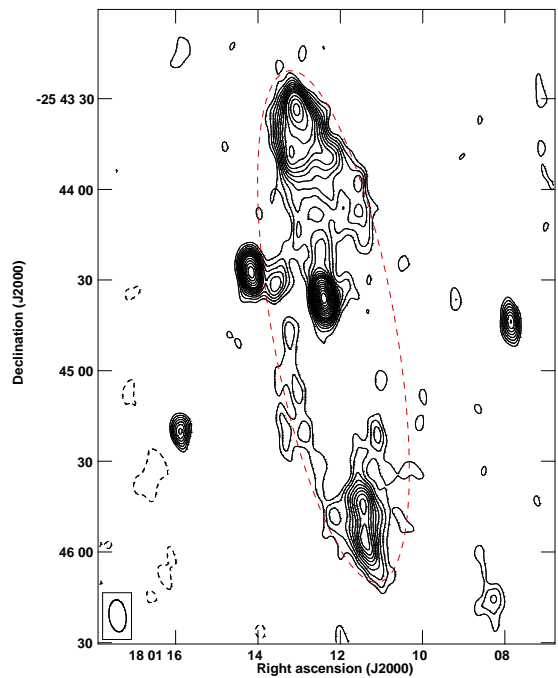}
      \caption{
      Natural-weight 6 cm map of \grs\ and its surrounding cocoon diffuse emission obtained by combining all VLA data sets (see  Table \ref{log}). 
      The horizontal bar in the top right corner gives the angular scale. 
      The dashed red ellipse sketches the edges of a previously unseen, 
      cocoon-like structure around this microquasar. The synthesized beam of 10\pri 6 $\times$ 5\pri 6, with position angle 6\grp 7, is displayed by the bottom left ellipse.
      Contours shown correspond to $-3$, 3, 4, 5, 6, 7, 8, 9, 10, 11 ,12, 14, 16, 18, 20, 22, 24, and 26 times  the rms background noise of 6 $\mu$Jy beam$^{-1}$. 
      Three unrelated compact radio sources also appear in this map.
              } 
         \label{cocoon}
   \end{figure}

We emphasize that changes in the microquasar central engine cannot be responsible for the morphological disruption of the outer jet, 
since the cooling time $t_c$ of relativistic electrons by synchrotron radiation in the jet head is too long. 
Specifically, $t_c[{\rm s}] \sim  5 \times 10^8 \gamma^{-1} B^{-2}$, where $\gamma$ is the Lorentz factor of the electrons and $B$ the magnetic field in Gauss.
For $B \sim 10^{-5}$ G and $\gamma \sim 10^6$, we get $t_c \sim 3$ Myr! 
It is clear that the structure must be destroyed and the electrons then diffuse into the interstellar medium, 
escaping from the region where they were confined. This plasma leaving the hotspot can form a cocoon structure around the radio lobes, 
creating cavities or bubbles, as seen in the environment of Galactic and extragalactic sources of relativistic jets 
with continued activity  \citep{2006MNRAS.370.1513K, 2005Natur.436..819G,  2006ApJ...644L...9W, 2010Natur.466..209P}.
In this context, it makes sense to wonder if this cocoon structure also exists in \grs.  To shed light on this issue, 
we decided to concatenate all observational projects in  Table \ref{log} to produce the deepest radio map available for this microquasar (see Fig. \ref{cocoon}). 
The total on-source integration time of this huge data set amounts to about $19^h$,  allowing us to reach a background rms noise of 6  $\mu$Jy beam$^{-1}$  with natural weight. 
In this map, bridges of extended emission emanate from both radio lobes and almost surround the whole bipolar jet complex with an elliptical shape. 
The faintest cocoon edges appear at a $4\sigma$ level and cover more than 50\% of the elliptical perimeter.
Such diffuse features reveal, for the first time, the expected cocoon-like structure around \grs. 
This is a new similarity between microquasars and radio galaxies that has never been observed before.


New, more sensitive multi-wavelength observations will provide a benchmark for validating our current ideas of the processes involving collimated flows in
astrophysical contexts, which otherwise could not be appropriately sampled in time. For instance, detailed changes
in jet structures and fainter cocoon shells created by jet relict particles should be revealed by new  radio interferometers,  such as EVLA, LOFAR, or SKA.



 
 

\begin{acknowledgements}

This work was supported by grant AYA2013-47447-C3-3-P from the Spanish Ministerio de Econom\'{\i}a y Competitividad (MINECO), 
and by the Consejer\'{\i}a de Econom\'{\i}a, Innovaci\'on, Ciencia y Empleo of Junta de Andaluc\'{\i}a under excellence grant FQM-1343 and research group FQM-322, as well as FEDER funds. GER is a member of CONICET. The National Radio Astronomy Observatory is a facility of the National Science Foundation operated under cooperative agreement by Associated Universities, Inc. 

\end{acknowledgements}



\bibliographystyle{aa} 
\bibliography{references} 
     

\end{document}